\documentclass[conference]{IEEEtran}
\IEEEoverridecommandlockouts
\usepackage{cite}
\usepackage{csquotes}
\usepackage{amsmath,amssymb,amsfonts}
\usepackage{algorithmic}
\usepackage{graphicx}
\usepackage{textcomp}
\usepackage{xcolor}
\usepackage{xspace}
\usepackage[hyphens]{url}
\usepackage{hyperref}
\usepackage[hyphenbreaks]{breakurl}
\usepackage[keeplastbox]{flushend}
\usepackage{scalerel}
\usepackage{tikz}
\usetikzlibrary{svg.path}

\newcommand{\AI}{\textsc{AI}\xspace}
\newcommand{\ICT}{\textsc{ICT}\xspace}
\newcommand{\GDPR}{\textsc{GDPR}\xspace}
\newcommand{\ADM}{\textsc{ADM}\xspace}

\newcommand\copyrighttext{%
  \footnotesize \textcopyright~2021 IEEE. Personal use of this material is permitted.
  Permission from IEEE must be obtained for all other uses, in any current or future
  media, including reprinting/republishing this material for advertising or promotional
  purposes, creating new collective works, for resale or redistribution to servers or
  lists, or reuse of any copyrighted component of this work in other works.
  }
\newcommand\copyrightnotice{%
\begin{tikzpicture}[remember picture,overlay]
\node[anchor=south,yshift=20pt] at (current page.south) {\fbox{\parbox{\dimexpr\textwidth-\fboxsep-\fboxrule\relax}{\copyrighttext}}};
\end{tikzpicture}%
}

\begin{document}

\title{Explainability Auditing for Intelligent Systems: \\ A Rationale for Multi-Disciplinary Perspectives
}

\author{\IEEEauthorblockN{Markus Langer\IEEEauthorrefmark{1}\IEEEauthorrefmark{5},
Kevin Baum\IEEEauthorrefmark{1}\IEEEauthorrefmark{5},
Kathrin Hartmann\IEEEauthorrefmark{2}\IEEEauthorrefmark{5}, 
Stefan Hessel\IEEEauthorrefmark{3}\IEEEauthorrefmark{5},
Timo Speith\IEEEauthorrefmark{1}\IEEEauthorrefmark{5} and
Jonas Wahl\IEEEauthorrefmark{4}\IEEEauthorrefmark{5}}

\IEEEauthorblockA{\IEEEauthorrefmark{1}Saarland University, Saarbrücken, Germany\\ Email: \{markus.langer, kevin.baum, timo.speith\}@uni-saarland.de}

\IEEEauthorblockA{\IEEEauthorrefmark{2}TU Kaiserslautern, Kaiserslautern, Germany\\ Email: kathrin.hartmann@sowi.uni-kl.de}

\IEEEauthorblockA{\IEEEauthorrefmark{3}reuschlaw Legal Consultants, Saarbrücken, Germany\\ Email: stefan.hessel@reuschlaw.de}

\IEEEauthorblockA{\IEEEauthorrefmark{4}University of Bonn, Bonn, Germany\\ Email: wahl@iam.uni-bonn.de}

\IEEEauthorblockA{\IEEEauthorrefmark{5}Algoright e.V., Germany\\ Email: firstname.lastname@algoright.de}

}

\maketitle

\copyrightnotice
\vspace{-2ex}

\begin{abstract}
National and international guidelines for trustworthy artificial intelligence (\AI) consider explainability to be a central facet of trustworthy systems. This paper outlines a multi-disciplinary rationale for explainability auditing. Specifically, we propose that explainability auditing can ensure the quality of explainability of systems in applied contexts and can be the basis for certification as a means to communicate whether systems meet certain explainability standards and requirements. Moreover, we emphasize that explainability auditing needs to take a multi-disciplinary perspective, and we provide an overview of four perspectives (technical, psychological, ethical, legal) and their respective benefits with respect to explainability auditing.

\end{abstract}

\begin{IEEEkeywords}
Auditing, Certification, Explainability, Explainable Artificial Intelligence, Requirements, Trustworthy AI
\end{IEEEkeywords}

\section{Introduction}
National and international guidelines for trustworthy artificial intelligence (\AI) consider explainability (as well as related concepts such as transparency and interpretability) to be a central facet of trustworthy systems (i.e., systems that can be trusted) \cite{BMBF2018Strategie, EU2019Ethics}. In fact, explainability seems to be the most commonly featured concept in regulatory guidelines on \AI around the globe \cite{Jobin2019Global, Hagendorff2020Ethics}. Thus, ensuring system explainability seems to be a central step towards trustworthy \AI.

As it is considered to be a central enabler of many crucial desiderata associated with intelligent systems \cite{Chazette2021Exploring, Arrieta2020Explainable, Langer2021What}, the importance of explainability for overall system trustworthiness becomes even more apparent. These desiderata can take the form of goals, interests, needs, and demands of the multiple stakeholders involved in the development, deployment, and actual use of intelligent systems \cite{Langer2021What}. For example, such desiderata could be to have usable, robust, or accountable systems \cite{Arrieta2020Explainable, Burrell2016Machine, Langer2021What}. With regard to such desiderata, explainability aims to ensure an improved understanding of system processes and outputs to a) enable stakeholders to more easily decide whether a specific desideratum is fulfilled, and b) facilitate improvement of a system with respect to a given desideratum \cite{Langer2021What}. For instance, better understanding of system processes and outputs can help to improve system usability for users \cite{Chazette2020Explainability}, can aid developers to increase system robustness \cite{Carvalho2019Machine}, can help regulating bodies to clarify legal and ethical accountability in case of unfavorable outcomes \cite{Binns2018Reducing}, and can lead to more warranted trust in intelligent systems \cite{Langer2021What, Jacovi2021Formalizing}.

However, ensuring and assessing system explainability in applied contexts is challenging. For developers, it is demanding to design systems that provide insights into their decision processes and that enable other stakeholders to better understand these processes \cite{DoshiVelez2017Towards}. For users, it will be only after some experience with a system that they realize whether they can understand its decision processes and outputs \cite{ko2011profiting}. Similarly, regulating bodies might only realize after an unfavorable outcome has happened that they are not able to understand what kind of system malfunction has led to this outcome. Adding to this complexity, explainability is a multi-disciplinary area of research and practice \cite{Adadi2018Peeking}. Consequently, for a comprehensive picture regarding explainability in applied contexts, explainabiltiy needs to be analyzed from the perspectives of multiple disciplines \cite{Langer2021What}. For instance, focusing solely on the technical perspective of explainability (e.g., how to design systems that, in principle, can provide insights into their decision-making processes) will lack a human-centered analysis of how explainability can fulfill societal desiderata \cite{DoshiVelez2017Towards} For this, we might additionally need a psychological perspective which empirically investigates whether a specific explainability approach has the intended effects on human-system interaction. In addition, explainability is associated with ethical and legal questions (e.g., concerning the allocation of responsibility, or the General Data Protection Regulation of the European Union) and, thus, these perspectives complements the technical and psychological perspectives.

In this vision paper, we highlight that explainability auditing and certification, as a way to ensure and assess system explainability, needs to take multi-disciplinary perspectives to enable a comprehensive analysis of explainability in applied contexts. During audits, auditors investigate whether products or processes meet certain quality standards and requirements. Certification is a way to communicate that a product or process has undergone quality control or suffices certain requirements, and can thus be an outcome of auditing processes. Both auditing and certification can help stakeholders to quickly assess whether products or processes follow certain quality standards and, in consequence, adequately evaluate their alignment with respect to these standards. However, explainability auditing has only received brief notion in previous work (e.g., in \cite{Moekander2021Ethics}). In this paper we reinforce this idea and, moreover, highlight that explainability auditing is only possible by taking a multi-disciplinary perspective on the auditing process, as no single discipline can succeed in fully capturing the complexity of defining requirements on explainability in systems. 

This paper is structured along four different perspectives for explainability auditing (i.e., technical, psychological, legal, and ethical). For each of these perspectives, we present dimensions that might be necessary to consider in auditing processes. Additionally, we present possible benefits that may result from ensuring that explainability meets these dimensions.

\section{Why Explainability Auditing?}
In what follows, we emphasize different perspectives on explainability auditing that provide a rationale regarding a) what dimensions we need to investigate in an explainability auditing process, and b) what benefits we can expect when these dimensions are met. Clearly, this is not a comprehensive list of perspectives on explainability auditing, as we could imagine further perspectives to complement our multi-disciplinary approach (e.g., a sociological perspective). However, this list is intended to envision how different perspectives may come together in order to ensure system explainability.


\subsection{Technical Perspective}

\subsubsection{Technical Dimensions}
From a technical perspective, explainability auditing needs to assess the current status of a system's explainability. Following, we present sample auditing dimensions from a technical perspective (for a more extensive list of technical auditing dimensions, see \cite{Sokol2020Explainability}).

\paragraph{Functional explainability} Is a system designed in such a way that it allows for human insight, or does it provide additional methods that shed light onto its decision-processes? In the first case, we call a system ante-hoc, in the second-case post-hoc explainable \cite{Arrieta2020Explainable}. Testing whether a system is capable of providing intelligibility to humans is a basic requirement for system explainability. This may happen on two levels: either the system is explainable with regards to specific outputs (local explainability), or the system's decision-making process is explainable as a whole (global explainability) \cite{Sokol2020Explainability, Hall2019Systematic}.

\paragraph{Faithful explainability} Does the system provide information that describe its decision-making process accurately and truthfully? Ensuring that a system provides faithful insights into its decision-making processes is essential for its trustworthiness and the information's reliability \cite{Rosenfeld2019Explainability}. For instance, faithful explainability is likely present in systems that reveal actual causal chains of their decision processes \cite{Miller2019Explanation}.
 
\paragraph{Interactive explainability} Can the system's explanations adapt or be adapted to respective stakeholder needs? Most contemporary explainability approaches rely on a one-size-fits-all solution when delivering explanations \cite{Sokol2020Explainability}. However, to provide useful information to stakeholders, explainability approaches need to integrate stakeholders' background knowledge and their explanatory needs given a particular system. Interactive explainability seems to be key with respect to these requirements \cite{Sokol2020Explainability, Schneider2019Personalized}.

\paragraph{Explainability trade-off} Is it more important to have an explainable system, or should the system rather be efficient or accurate? Some intelligent systems are or can be made explainable (at least to some degree) without a loss in efficiency or accuracy, but for others this is not technically feasible \cite{Burrell2016Machine, Sokol2020Explainability}. Whether to trade-off accuracy for explainability is a decision that must not only be made, but must also be justified, and should, therefore, be included in an auditing process.

\subsubsection{Technical Benefits}
Explainability can be an important building block for better systems. Primarily, explainability can help developers to detect errors and, thus, can lead to increased system debugability, facilitating system safety and robustness \cite{Carvalho2019Machine}. Further benefits are verification and validation, which become easier through explainability \cite{Darlington2013Aspects}. Overall, the technical perspective is a foundation for all other perspectives on explainability because without the technical perspective, requirements from other perspective cannot be met.


\subsection{Psychological Perspective}

\subsubsection{Psychological Dimensions}
The psychological perspective on system explainability mostly reflects user needs in the respective application context of an intelligent system. Following, we present sample auditing dimensions from a psychological perspective.

\paragraph{Understandability} Does the provided information help people to better understand system decision-making processes? This may be the primary psychological dimension to investigate in explainability audits \cite{Langer2021What, Koehl2019Explainability} as much of the previous work has resulted in explainability approaches aimed towards helping developers understand system decision processes but not other stakeholders \cite{Burrell2016Machine, Paez2019Pragmatic}. 

\paragraph{Context-Dependency} Does the system provide con\-text-related intelligibility of its decision-making processes or of its outputs? In this case, context-related means that people need insights that depend on their goals and needs relevant to the context \cite{Langer2021What, Sokol2020Explainability}. For instance, if medical doctors want to learn why a system produced a respective diagnosis, they may want to have detailed information helping them to understand why a system produced a respective output. In contrast, if they are under time-pressure and want to quickly decide what might be the best patient treatment, different kind of information might be more helpful \cite{Ackerman2012Taking}.

\paragraph{Usability} Does the system provide easy to access information, and are explainability functionalities easy to use \cite{Venkatesh2003User, Chromik2019Dark}? System explainability needs to be usable, meaning that people can actually use the system in a way that they can access the information they need to better understand the system's decision-making processes. Ensuring usable explainability can mean to optimize user interfaces or interactivity between user and system \cite{Miller2019Explanation}.

\paragraph{Honesty} Does the system provide non-deceptive information? There are emerging discussions on possible \enquote{dark patterns} of explainability \cite{Chromik2019Dark}. Explainability auditing needs to explore whether system explainability contributes to the goals of respective stakeholders or whether the system is designed to nudge or persuade people instead of providing actually relevant information. In the case of honest explainability, stakeholder interests might diverge as system deployers might intentionally want systems to be designed in a way that ensures certain user behavior whereas users might be less happy about systems that try to influence their behavior \cite{Burrell2016Machine}. 

\subsubsection{Psychological Benefits}
First, ensuring the psychological dimensions may help to develop warranted trust in systems \cite{Jacovi2021Formalizing, Lee2004trust}. If we ensure that system explainability meets the psychological dimensions outlined above, people will be better able to assess when and under what conditions to follow system recommendations. 
Second, ensuring psychological dimensions of explainability can increase system acceptance and thus actual use of systems in applied contexts \cite{Venkatesh2003User}. Although maybe obvious, designing explainability qualities of systems in a way that are relevant and usable will ensure that systems will actually end up being used instead of being ignored.
Third, joint human-system performance can improve \cite{Lai2019human}. If systems provide understandable, context-dependent, usable, and honest information, this might enable users to make more informed and more accurate decisions in contrast to situations where they would receive too complex, irrelevant, or deceptive information for their respective task at hand. 


\subsection{Legal Perspective}

\subsubsection{Legal Dimensions}
From a legal perspective, the auditing process should take into account all legal requirements for the use of intelligent systems, in order to be able to prove to the user that the intelligent system operates in compliance with the law (especially with regard to data protection and cybersecurity) and, in particular, that no fundamental rights of the individual are violated. Intelligent systems that are not able to prove their compliance with the legal system (as minimum requirement) cannot be considered trustworthy and explainable. In the future, the auditing process will also have to take into account regulatory requirements for the use of intelligent systems, such as those set out in the \textsc{EU} Commission's Proposal for a Regulation laying down harmonised rules on artificial intelligence (\textsc{COM}(2021) 206 final).

\paragraph{General Data Protection Regulation (\GDPR)}
When is the processing of data permissible under data protection law when using intelligent systems? Insofar as personal data is processed via an intelligent system, the requirements of the \GDPR must be observed. These oblige the controller (\emph{Art. 4 No. 7 \GDPR}) to comply with the principles relating to processing of personal data (\emph{Art. 5 \GDPR}) as well as more detailed requirements, such as proof of a sufficient legal basis for data processing (\emph{Art. 6 \GDPR}) as well as measures for security (\emph{Art. 32 (1) \GDPR}). In addition to the general data protection requirements, \emph{Art. 22 \GDPR} may also need to be observed, which sets out requirements for automated decision-making (\ADM). However, \emph{Art. 22 \GDPR} only applies if the \ADM decision is not reviewed again by a human but directly translated into a decision of its own \cite{SWB-1651766754}. Thus, if a human correctness and plausibility check of the decision takes place, \emph{Art. 22 \GDPR} does not make any additional specifications \cite{SWB-1700373293}. However, if it is a decision based solely on automated processing that produces legal effects concerning the data subject or significantly affects them, it is only permissible if the requirements of \emph{Art. 22 (2) \GDPR} are fulfilled. This is the case if the decision is necessary for a contract between the data subject and the controller, if it is required by law, or if it is based on the data subject's explicit consent.

\paragraph{Cybersecurity Act}
What are the legal requirements for cybersecurity? In addition to the objectives, tasks and organizational matters of the European Union Agency for Cybersecurity (\textsc{ENISA}), the Cybersecurity Act also contains a framework for the establishment of a European cybersecurity certification for information and communications technology (\ICT) products, \ICT services and \ICT processes, according to \emph{Art. 1 (1) of the Cybersecurity Act}. According to Art. 2 No. 13 Cybersecurity Act, \ICT services refer to services that consist entirely or predominantly of the transmission, storage, retrieval or processing of information by means of network and information systems. According to \emph{Art. 2 No. 14 Cybersecurity Act}, the term \ICT process includes all activities to design, develop, provide or maintain an \ICT product or service. Intelligent systems, since they administer information by means of network and information systems, fulfill the requirements of an \ICT service. In addition, activities related to intelligent systems may also meet the requirements for an \ICT process if the aforementioned requirements are met.

However, the Cybersecurity Act does not impose any mandatory cybersecurity requirements on intelligent systems. \emph{Art. 46 et seq. of the Cybersecurity Act} merely provide for a voluntary certification framework. Therefore, there is no obligation for manufacturers or operators to carry out certification, meaning that no binding requirements for the cybersecurity of intelligent systems have yet resulted from the Cybersecurity Act. At the same time, the lack of specific cybersecurity requirements on intelligent systems shows that current regulation does not sufficiently address new technologies and their particular threats. A legal framework that proactively shapes the use of intelligent systems would therefore be desirable.

\paragraph{Proposal for a regulation laying down harmonised rules on artificial intelligence (Artificial Intelligence Act)}
Are there legal requirements for the use of intelligent systems? On April 21, 2021, the \textsc{EU} Commission published its proposal for a regulation laying down harmonised rules on artificial intelligence (COM(2021) 206 final). The planned regulation has clear parallels to the \GDPR in several places, e.g. the risk-based approach and the scope of application. For example, the regulation applies not only to providers in the \textsc{EU}, but also to providers who offer their \AI systems on the European market (market location principle). \emph{Art. 5 of the planned regulation} prohibits certain use-case scenarios for \AI, e.g. discrimination. In addition, \emph{Art. 6 et seq. of the planned regulation} specifies high-risk areas of application for which stricter requirements apply than for ordinary \AI applications. The Artificial Intelligence Act is expected to have an enormous impact on the legal framework for the use of intelligent systems in Europe, but also worldwide. As soon as the regulation is available in its final version, it should therefore be included in explainability audits and later be a mandatory part of the auditing process.

\subsubsection{Legal Benefits}
From a legal point of view, an auditing process promises the advantage that compliance with legal obligations can be achieved and demonstrated to third parties. In particular, if auditing is carried out by an independent third party, the process creates the opportunity to actively enforce the law by making the auditing itself a legal obligation or by attaching legal benefits to it. This can also increase the incentive to audit an intelligent system.


\subsection{Ethical Perspective}

\subsubsection{Ethical Dimensions}
The ethical perspective on system explainability is build on two fundamental considerations. First, it is about upholding moral rights (e.g., as given by certain normative theories) of and fulfilling norms with respect to the involved stakeholders. Second, it is about enabling stakeholders to live up to their obligations. Following, we present sample auditing dimensions from an ethical perspective.

\paragraph{Responsibility} Does the provided information enable responsible decision-making? And: Does the information make the allocation of (moral) responsibility possible? Responsibility is about identifying who is blameworthy or answerable for a certain decision of a system \cite{Strawson1962Freedom}. Especially in cases where a human is in the loop and has to make a decision based on a system's output, it is important to consider both of the above facets when incorporating explainability into a system. First, the provision of certain pieces of information may enable a human in the loop to become a responsible decision maker \cite{Robbins2019Misdirected}. Subsequently, it becomes possible to allocate responsibility to this person \cite{Pieters2011Explanation}. However, not all types of explainability may help to achieve responsibility \cite{Langer2021What}.

\paragraph{Non-Discrimination} Does the system provide information that makes it possible to detect or at least check for potential, and assess actual discrimination? Especially for people affected by decisions, the decisions of intelligent systems can have significant implications. Be it applying for a loan, a job, or a visa, in all these cases intelligent systems are increasingly used. However, system outputs that lead to decisions about people may involve unfair biases (see, e.g., \cite{Lepri2018Fair}). The explanation of a rejection made by an intelligent system should make it possible to identify whether protected attributes like race or gender (also indirectly) affected the decision. Furthermore, where the influence of protected attributes cannot be prevented, explanations should at least enable parties affected by automated system-based decisions to understand why system outputs are biased and whether this bias is tolerable or even justified \cite{Kusner2017Counterfactual}.

\paragraph{(Moral) Right to Explanation} Does the information provision comply with moral rights to explanation? Arguably, there is a moral right to explanation that requires intelligent systems to be able to provide certain types of explanations \cite{Kim2021Why}. More precisely, people should receive explanations that enable them to contest decisions that are based on the recommendations of intelligent systems \cite{Tubella2020Contestable}. In general, the advance of intelligent systems to ever more areas of human lives often precludes people affected by system-based decisions from tracing how certain decisions came about and affected them \cite{MoraCantallops2021Traceability}. This lack of traceability is morally problematic \cite{Baum2018From, Baum2018Towards}. Some types of explanations promise to empower people to contest and check decisions of intelligent systems \cite{Wachter2018Counterfactual}.

\subsubsection{Ethical Benefits}
The moral integrity of a system is a significant building block of its trustworthiness. As such, the possibility to check for this integrity, for instance, by checking whether the system unduly discriminates, is essential for stakeholders to develop appropriate trust into systems. Furthermore, the possibility to check for such an integrity can also contribute to system trustworthiness (at least when viewing trustworthiness from a philosophical point of view \cite{McLeod2020Trust}). Lastly, systems that allow for 1) responsible decision-making, 2) an adequate allocation of responsibility, and 3) general contesting of decisions (as given by a moral right for explanations) are important in ensuring acceptance of decisions of intelligent systems \cite{Baum2018From, Rosenfeld2019Explainability}.

\section{Outlook and Conclusion}
The main contribution of this vision paper is to propose multi-disciplinary dimensions for explainability auditing. While explainability auditing based on only one of the aforementioned perspectives may provide valuable quality control for this perspective (e.g., with respect to regulatory requirements), it will ultimately fall short with respect to the multi-disciplinary challenges associated with explainability.

The dimensions we have proposed in this paper are just a rough description of what could be the basis for explainability auditing. Future work would need to outline concrete auditing and certification processes and investigate what might be the most practical implementation of explainability auditing. In this sense, the proposed dimensions could serve as a starting point for creating an auditing checklist. To this end, future research might need to develop comprehensive lists of dimensions that are relevant in explainability auditing processes \cite{Langer2021What}. 

Auditing processes can be a first step towards quality norms (e.g., \textsc{DIN} norms) as well as a way to ensure these norms as soon as they are implemented. We hope that this paper initiates a discussion on the ways in which multi-disciplinary explainability auditing processes can be realized in practice.

\section*{Acknowledgments}
Work on this paper was funded by the Volkswagen Foundation grants \textsc{AZ 98512}, 98513, and 98514 \href{https://explainable-intelligent.systems}{\enquote{Explainable Intelligent Systems}} (\textsc{EIS}) and by the \textsc{DFG} grant 389792660 as part of \href{https://perspicuous-computing.science}{\textsc{TRR}~248}. We thank three anonymous reviewers for their feedback.

\bibliographystyle{IEEEtran}
\bibliography{bibliography}

\end{document}